
%
%
%
%
%
%
%
%
\def\standardrisposta{s }\def\reducedrisposta{r }
\def\mplarisposta{mpla }\def\zerorisposta{z }
\def\doublerisposta{d }\def\cartarisposta{e }\def\amsrisposta{y }
\newcount\ingrandimento \newcount\sinnota \newcount\dimnota
\newcount\unoduecol \newdimen\collhsize \newdimen\tothsize
\newdimen\fullhsize \newcount\controllorisposta \sinnota=1
\newskip\infralinea  \global\controllorisposta=0
\immediate\write16 { ********  Welcome to PANDA macros (Plain TeX,
AP, 1991) ******** }
\immediate\write16 { You'll have to answer a few questions in
lowercase.}
\message{>  Do you want it in double-page (d), reduced (r)
or standard format (s) ? }\read-1 to\risposta
\message{>  Do you want it in USA A4 (u) or EUROPEAN A4
(e) paper size ? }\read-1 to\srisposta
\message{>  Do you have AMSFonts 2.0 (math) fonts (y/n) ? }
\read-1 to\arisposta
%
%
%
%
%
\ifx\risposta\standardrisposta \ingrandimento=1200
\message {>> This will come out UNREDUCED << }
\dimnota=2 \unoduecol=1 \global\controllorisposta=1 \fi
\ifx\risposta\reducedrisposta \ingrandimento=1095 \dimnota=1
\unoduecol=1  \global\controllorisposta=1
\message {>> This will come out REDUCED << } \fi
\ifx\risposta\doublerisposta \ingrandimento=1000 \dimnota=2
\unoduecol=2   \message {>> You must print this in
LANDSCAPE orientation << } \global\controllorisposta=1 \fi
\ifx\risposta\mplarisposta \ingrandimento=1000 \dimnota=1
\message {>> Mod. Phys. Lett. A format << }
\unoduecol=1 \global\controllorisposta=1 \fi
\ifx\risposta\zerorisposta \ingrandimento=1000 \dimnota=2
\message {>> Zero Magnification format << }
\unoduecol=1 \global\controllorisposta=1 \fi
\ifnum\controllorisposta=0  \ingrandimento=1200
\message {>>> ERROR IN INPUT, I ASSUME STANDARD
UNREDUCED FORMAT <<< }  \dimnota=2 \unoduecol=1 \fi
\magnification=\ingrandimento
%
%
%
%
\newdimen\eucolumnsize \newdimen\eudoublehsize \newdimen\eudoublevsize
\newdimen\uscolumnsize \newdimen\usdoublehsize \newdimen\usdoublevsize
\newdimen\eusinglehsize \newdimen\eusinglevsize \newdimen\ussinglehsize
\newskip\standardbaselineskip \newdimen\ussinglevsize
\newskip\reducedbaselineskip \newskip\doublebaselineskip
\eucolumnsize=12.0truecm    
\eudoublehsize=25.5truecm   
\eudoublevsize=6.5truein    
\uscolumnsize=4.4truein     
\usdoublehsize=9.4truein    
\usdoublevsize=6.8truein    
\eusinglehsize=6.5truein    
\eusinglevsize=24truecm     
\ussinglehsize=6.5truein    
\ussinglevsize=8.9truein    
\standardbaselineskip=16pt plus.2pt  
\reducedbaselineskip=14pt plus.2pt   
\doublebaselineskip=12pt plus.2pt    
%
%
\def\Portoffset{}
\def\Landoffset{}
\ifx\risposta\mplarisposta \def\Portoffset{\hoffset=1.8truecm} \fi
%
%
\def\Landspec{}
\tolerance=10000
\parskip=0pt plus2pt  \leftskip=0pt \rightskip=0pt
%
%
\ifx\risposta\standardrisposta \infralinea=\standardbaselineskip \fi
\ifx\risposta\reducedrisposta  \infralinea=\reducedbaselineskip \fi
\ifx\risposta\doublerisposta   \infralinea=\doublebaselineskip \fi
\ifx\risposta\mplarisposta     \infralinea=13pt \fi
\ifx\risposta\zerorisposta     \infralinea=12pt plus.2pt\fi
\ifnum\controllorisposta=0    \infralinea=\standardbaselineskip \fi
\ifx\risposta\doublerisposta   \Landoffset \else \Portoffset \fi
\ifx\risposta\doublerisposta \ifx\srisposta\cartarisposta
\tothsize=\eudoublehsize \collhsize=\eucolumnsize
\vsize=\eudoublevsize  \else  \tothsize=\usdoublehsize
\collhsize=\uscolumnsize \vsize=\usdoublevsize \fi \else
\ifx\srisposta\cartarisposta \tothsize=\eusinglehsize
\vsize=\eusinglevsize \else  \tothsize=\ussinglehsize
\vsize=\ussinglevsize \fi \collhsize=4.4truein \fi
\ifx\risposta\mplarisposta \tothsize=5.0truein
\vsize=7.8truein \collhsize=4.4truein \fi
%
%
%
%
\newcount\contaeuler \newcount\contacyrill \newcount\contaams
\font\ninerm=cmr9  \font\eightrm=cmr8  \font\sixrm=cmr6
\font\ninei=cmmi9  \font\eighti=cmmi8  \font\sixi=cmmi6
\font\ninesy=cmsy9  \font\eightsy=cmsy8  \font\sixsy=cmsy6
\font\ninebf=cmbx9  \font\eightbf=cmbx8  \font\sixbf=cmbx6
\font\ninett=cmtt9  \font\eighttt=cmtt8  \font\nineit=cmti9
\font\eightit=cmti8 \font\ninesl=cmsl9  \font\eightsl=cmsl8
\skewchar\ninei='177 \skewchar\eighti='177 \skewchar\sixi='177
\skewchar\ninesy='60 \skewchar\eightsy='60 \skewchar\sixsy='60
\hyphenchar\ninett=-1 \hyphenchar\eighttt=-1 \hyphenchar\tentt=-1
%
\font\tencmmib=cmmib10  \newfam\cmmibfam  \skewchar\tencmmib='177
\font\tencmbsy=cmbsy10  \newfam\cmbsyfam  \skewchar\tencmbsy='60
\def\scaps{\cmcsc}                 
\font\tencmcsc=cmcsc10  \newfam\cmcscfam
\ifnum\ingrandimento=1095

\font\capsone=cmcsc10 at 10.95pt 

\else

\font\capsone=cmcsc10 at 12pt 
\fi

\def\ttaarr{\bf}		
\def\ppaarr{\sl}		

%
%
%
\newfam\eufmfam \newfam\msamfam \newfam\msbmfam \newfam\eufbfam
\def\Loadeulerfonts{\global\contaeuler=1 \ifx\arisposta\amsrisposta
\font\teneufm=eufm10              
\font\eighteufm=eufm8 \font\nineeufm=eufm9 \font\sixeufm=eufm6
\font\seveneufm=eufm7  \font\fiveeufm=eufm5
\font\teneufb=eufb10              
\font\eighteufb=eufb8 \font\nineeufb=eufb9 \font\sixeufb=eufb6
\font\seveneufb=eufb7  \font\fiveeufb=eufb5
\font\teneurm=eurm10              
\font\eighteurm=eurm8 \font\nineeurm=eurm9
\font\teneurb=eurb10              
\font\eighteurb=eurb8 \font\nineeurb=eurb9
\font\teneusm=eusm10              
\font\eighteusm=eusm8 \font\nineeusm=eusm9
\font\teneusb=eusb10              
\font\eighteusb=eusb8 \font\nineeusb=eusb9
\else \def\eufm{\tt} \def\eufb{\tt} \def\eurm{\tt} \def\eurb{\tt}
\def\eusm{\tt} \def\eusb{\tt}    \fi}

\def\loadamsmath{\global\contaams=1 \ifx\arisposta\amsrisposta
\font\tenmsam=msam10 \font\ninemsam=msam9 \font\eightmsam=msam8
\font\sevenmsam=msam7 \font\sixmsam=msam6 \font\fivemsam=msam5
\font\tenmsbm=msbm10 \font\ninemsbm=msbm9 \font\eightmsbm=msbm8
\font\sevenmsbm=msbm7 \font\sixmsbm=msbm6 \font\fivemsbm=msbm5
\else \def\msbm{\bf} \fi \def\Bbb{\msbm} \def\symbl{\msam} \tenpoint}
\def\loadcyrill{\global\contacyrill=1 \ifx\arisposta\amsrisposta
\font\tenwncyr=wncyr10 \font\ninewncyr=wncyr9 \font\eightwncyr=wncyr8
\font\tenwncyb=wncyr10 \font\ninewncyb=wncyr9 \font\eightwncyb=wncyr8
\font\tenwncyi=wncyr10 \font\ninewncyi=wncyr9 \font\eightwncyi=wncyr8
\else \def\cyrill{\sl} \def\cyrilb{\sl} \def\cyrili{\sl} \fi\tenpoint}
\ifx\arisposta\amsrisposta
\font\sevenex=cmex7               
\font\eightex=cmex8  \font\nineex=cmex9
\font\ninecmmib=cmmib9   \font\eightcmmib=cmmib8
\font\sevencmmib=cmmib7 \font\sixcmmib=cmmib6
\font\fivecmmib=cmmib5   \skewchar\ninecmmib='177
\skewchar\eightcmmib='177  \skewchar\sevencmmib='177
\skewchar\sixcmmib='177   \skewchar\fivecmmib='177
\font\ninecmbsy=cmbsy9    \font\eightcmbsy=cmbsy8
\font\sevencmbsy=cmbsy7  \font\sixcmbsy=cmbsy6
\font\fivecmbsy=cmbsy5   \skewchar\ninecmbsy='60
\skewchar\eightcmbsy='60  \skewchar\sevencmbsy='60
\skewchar\sixcmbsy='60    \skewchar\fivecmbsy='60
\font\ninecmcsc=cmcsc9    \font\eightcmcsc=cmcsc8     \else
\def\cmmib{\fam\cmmibfam\tencmmib}\textfont\cmmibfam=\tencmmib
\scriptfont\cmmibfam=\tencmmib \scriptscriptfont\cmmibfam=\tencmmib
\def\cmbsy{\fam\cmbsyfam\tencmbsy} \textfont\cmbsyfam=\tencmbsy
\scriptfont\cmbsyfam=\tencmbsy \scriptscriptfont\cmbsyfam=\tencmbsy
\scriptfont\cmcscfam=\tencmcsc \scriptscriptfont\cmcscfam=\tencmcsc
\def\cmcsc{\fam\cmcscfam\tencmcsc} \textfont\cmcscfam=\tencmcsc \fi
\catcode`@=11
\newskip\ttglue
\gdef\tenpoint{\def\rm{\fam0\tenrm}
  \textfont0=\tenrm \scriptfont0=\sevenrm \scriptscriptfont0=\fiverm
  \textfont1=\teni \scriptfont1=\seveni \scriptscriptfont1=\fivei
  \textfont2=\tensy \scriptfont2=\sevensy \scriptscriptfont2=\fivesy
  \textfont3=\tenex \scriptfont3=\tenex \scriptscriptfont3=\tenex
  \def\mcal{\fam2 \tensy}  \def\mmit{\fam1 \teni}
  \textfont\itfam=\tenit \def\it{\fam\itfam\tenit}
  \textfont\slfam=\tensl \def\sl{\fam\slfam\tensl}
  \textfont\ttfam=\tentt \scriptfont\ttfam=\eighttt
  \scriptscriptfont\ttfam=\eighttt  \def\tt{\fam\ttfam\tentt}
  \textfont\bffam=\tenbf \scriptfont\bffam=\sevenbf
  \scriptscriptfont\bffam=\fivebf \def\bf{\fam\bffam\tenbf}
     \ifx\arisposta\amsrisposta    \ifnum\contaeuler=1
  \textfont\eufmfam=\teneufm \scriptfont\eufmfam=\seveneufm
  \scriptscriptfont\eufmfam=\fiveeufm \def\eufm{\fam\eufmfam\teneufm}
  \textfont\eufbfam=\teneufb \scriptfont\eufbfam=\seveneufb
  \scriptscriptfont\eufbfam=\fiveeufb \def\eufb{\fam\eufbfam\teneufb}
  \def\eurm{\teneurm} \def\eurb{\teneurb} \def\eusm{\teneusm}
  \def\eusb{\teneusb}    \fi    \ifnum\contaams=1
  \textfont\msamfam=\tenmsam \scriptfont\msamfam=\sevenmsam
  \scriptscriptfont\msamfam=\fivemsam \def\msam{\fam\msamfam\tenmsam}
  \textfont\msbmfam=\tenmsbm \scriptfont\msbmfam=\sevenmsbm
  \scriptscriptfont\msbmfam=\fivemsbm \def\msbm{\fam\msbmfam\tenmsbm}
     \fi      \ifnum\contacyrill=1     \def\cyrill{\tenwncyr}
  \def\cyrilb{\tenwncyb}  \def\cyrili{\tenwncyi}         \fi
  \textfont3=\tenex \scriptfont3=\sevenex \scriptscriptfont3=\sevenex
  \def\cmmib{\fam\cmmibfam\tencmmib} \scriptfont\cmmibfam=\sevencmmib
  \textfont\cmmibfam=\tencmmib  \scriptscriptfont\cmmibfam=\fivecmmib
  \def\cmbsy{\fam\cmbsyfam\tencmbsy} \scriptfont\cmbsyfam=\sevencmbsy
  \textfont\cmbsyfam=\tencmbsy  \scriptscriptfont\cmbsyfam=\fivecmbsy
  \def\cmcsc{\fam\cmcscfam\tencmcsc} \scriptfont\cmcscfam=\eightcmcsc
  \textfont\cmcscfam=\tencmcsc \scriptscriptfont\cmcscfam=\eightcmcsc
     \fi            \tt \ttglue=.5em plus.25em minus.15em
  \normalbaselineskip=12pt
  \setbox\strutbox=\hbox{\vrule height8.5pt depth3.5pt width0pt}
  \let\sc=\eightrm \let\big=\tenbig   \normalbaselines
  \baselineskip=\infralinea  \rm}
\gdef\ninepoint{\def\rm{\fam0\ninerm}
  \textfont0=\ninerm \scriptfont0=\sixrm \scriptscriptfont0=\fiverm
  \textfont1=\ninei \scriptfont1=\sixi \scriptscriptfont1=\fivei
  \textfont2=\ninesy \scriptfont2=\sixsy \scriptscriptfont2=\fivesy
  \textfont3=\tenex \scriptfont3=\tenex \scriptscriptfont3=\tenex
  \def\mcal{\fam2 \ninesy}  \def\mmit{\fam1 \ninei}
  \textfont\itfam=\nineit \def\it{\fam\itfam\nineit}
  \textfont\slfam=\ninesl \def\sl{\fam\slfam\ninesl}
  \textfont\ttfam=\ninett \scriptfont\ttfam=\eighttt
  \scriptscriptfont\ttfam=\eighttt \def\tt{\fam\ttfam\ninett}
  \textfont\bffam=\ninebf \scriptfont\bffam=\sixbf
  \scriptscriptfont\bffam=\fivebf \def\bf{\fam\bffam\ninebf}
     \ifx\arisposta\amsrisposta  \ifnum\contaeuler=1
  \textfont\eufmfam=\nineeufm \scriptfont\eufmfam=\sixeufm
  \scriptscriptfont\eufmfam=\fiveeufm \def\eufm{\fam\eufmfam\nineeufm}
  \textfont\eufbfam=\nineeufb \scriptfont\eufbfam=\sixeufb
  \scriptscriptfont\eufbfam=\fiveeufb \def\eufb{\fam\eufbfam\nineeufb}
  \def\eurm{\nineeurm} \def\eurb{\nineeurb} \def\eusm{\nineeusm}
  \def\eusb{\nineeusb}     \fi   \ifnum\contaams=1
  \textfont\msamfam=\ninemsam \scriptfont\msamfam=\sixmsam
  \scriptscriptfont\msamfam=\fivemsam \def\msam{\fam\msamfam\ninemsam}
  \textfont\msbmfam=\ninemsbm \scriptfont\msbmfam=\sixmsbm
  \scriptscriptfont\msbmfam=\fivemsbm \def\msbm{\fam\msbmfam\ninemsbm}
     \fi       \ifnum\contacyrill=1     \def\cyrill{\ninewncyr}
  \def\cyrilb{\ninewncyb}  \def\cyrili{\ninewncyi}         \fi
  \textfont3=\nineex \scriptfont3=\sevenex \scriptscriptfont3=\sevenex
  \def\cmmib{\fam\cmmibfam\ninecmmib}  \textfont\cmmibfam=\ninecmmib
  \scriptfont\cmmibfam=\sixcmmib \scriptscriptfont\cmmibfam=\fivecmmib
  \def\cmbsy{\fam\cmbsyfam\ninecmbsy}  \textfont\cmbsyfam=\ninecmbsy
  \scriptfont\cmbsyfam=\sixcmbsy \scriptscriptfont\cmbsyfam=\fivecmbsy
  \def\cmcsc{\fam\cmcscfam\ninecmcsc} \scriptfont\cmcscfam=\eightcmcsc
  \textfont\cmcscfam=\ninecmcsc \scriptscriptfont\cmcscfam=\eightcmcsc
     \fi            \tt \ttglue=.5em plus.25em minus.15em
  \normalbaselineskip=11pt
  \setbox\strutbox=\hbox{\vrule height8pt depth3pt width0pt}
  \let\sc=\sevenrm \let\big=\ninebig \normalbaselines\rm}
\gdef\eightpoint{\def\rm{\fam0\eightrm}
  \textfont0=\eightrm \scriptfont0=\sixrm \scriptscriptfont0=\fiverm
  \textfont1=\eighti \scriptfont1=\sixi \scriptscriptfont1=\fivei
  \textfont2=\eightsy \scriptfont2=\sixsy \scriptscriptfont2=\fivesy
  \textfont3=\tenex \scriptfont3=\tenex \scriptscriptfont3=\tenex
  \def\mcal{\fam2 \eightsy}  \def\mmit{\fam1 \eighti}
  \textfont\itfam=\eightit \def\it{\fam\itfam\eightit}
  \textfont\slfam=\eightsl \def\sl{\fam\slfam\eightsl}
  \textfont\ttfam=\eighttt \scriptfont\ttfam=\eighttt
  \scriptscriptfont\ttfam=\eighttt \def\tt{\fam\ttfam\eighttt}
  \textfont\bffam=\eightbf \scriptfont\bffam=\sixbf
  \scriptscriptfont\bffam=\fivebf \def\bf{\fam\bffam\eightbf}
     \ifx\arisposta\amsrisposta   \ifnum\contaeuler=1
  \textfont\eufmfam=\eighteufm \scriptfont\eufmfam=\sixeufm
  \scriptscriptfont\eufmfam=\fiveeufm \def\eufm{\fam\eufmfam\eighteufm}
  \textfont\eufbfam=\eighteufb \scriptfont\eufbfam=\sixeufb
  \scriptscriptfont\eufbfam=\fiveeufb \def\eufb{\fam\eufbfam\eighteufb}
  \def\eurm{\eighteurm} \def\eurb{\eighteurb} \def\eusm{\eighteusm}
  \def\eusb{\eighteusb}       \fi    \ifnum\contaams=1
  \textfont\msamfam=\eightmsam \scriptfont\msamfam=\sixmsam
  \scriptscriptfont\msamfam=\fivemsam \def\msam{\fam\msamfam\eightmsam}
  \textfont\msbmfam=\eightmsbm \scriptfont\msbmfam=\sixmsbm
  \scriptscriptfont\msbmfam=\fivemsbm \def\msbm{\fam\msbmfam\eightmsbm}
     \fi       \ifnum\contacyrill=1     \def\cyrill{\eightwncyr}
  \def\cyrilb{\eightwncyb}  \def\cyrili{\eightwncyi}         \fi
  \textfont3=\eightex \scriptfont3=\sevenex \scriptscriptfont3=\sevenex
  \def\cmmib{\fam\cmmibfam\eightcmmib}  \textfont\cmmibfam=\eightcmmib
  \scriptfont\cmmibfam=\sixcmmib \scriptscriptfont\cmmibfam=\fivecmmib
  \def\cmbsy{\fam\cmbsyfam\eightcmbsy}  \textfont\cmbsyfam=\eightcmbsy
  \scriptfont\cmbsyfam=\sixcmbsy \scriptscriptfont\cmbsyfam=\fivecmbsy
  \def\cmcsc{\fam\cmcscfam\eightcmcsc} \scriptfont\cmcscfam=\eightcmcsc
  \textfont\cmcscfam=\eightcmcsc \scriptscriptfont\cmcscfam=\eightcmcsc
     \fi             \tt \ttglue=.5em plus.25em minus.15em
  \normalbaselineskip=9pt
  \setbox\strutbox=\hbox{\vrule height7pt depth2pt width0pt}
  \let\sc=\sixrm \let\big=\eightbig \normalbaselines\rm }
\gdef\tenbig#1{{\hbox{$\left#1\vbox to8.5pt{}\right.\n@space$}}}
\gdef\ninebig#1{{\hbox{$\textfont0=\tenrm\textfont2=\tensy
   \left#1\vbox to7.25pt{}\right.\n@space$}}}
\gdef\eightbig#1{{\hbox{$\textfont0=\ninerm\textfont2=\ninesy
   \left#1\vbox to6.5pt{}\right.\n@space$}}}
\def\alternativefont#1#2{\ifx\arisposta\amsrisposta \relax \else
\xdef#1{#2} \fi}
\global\contaeuler=0 \global\contacyrill=0 \global\contaams=0
%
%
%
%
\newbox\fotlinebb \newbox\hedlinebb \newbox\leftcolumn
\gdef\makeheadline{\vbox to 0pt{\vskip-22.5pt
     \fullline{\vbox to8.5pt{}\the\headline}\vss}\nointerlineskip}
\gdef\makehedlinebb{\vbox to 0pt{\vskip-22.5pt
     \fullline{\vbox to8.5pt{}\copy\hedlinebb\hfil
     \line{\hfill\the\headline\hfill}}\vss} \nointerlineskip}
\gdef\makefootline{\baselineskip=24pt \fullline{\the\footline}}
\gdef\makefotlinebb{\baselineskip=24pt
    \fullline{\copy\fotlinebb\hfil\line{\hfill\the\footline\hfill}}}
\gdef\doubleformat{\shipout\vbox{\Landspec\makehedlinebb
     \fullline{\box\leftcolumn\hfil\columnbox}\makefotlinebb}
     \advancepageno}
\gdef\columnbox{\leftline{\pagebody}}
\gdef\line#1{\hbox to\hsize{\hskip\leftskip#1\hskip\rightskip}}
\gdef\fullline#1{\hbox to\fullhsize{\hskip\leftskip{#1}%
\hskip\rightskip}}
\gdef\footnote#1{\let\@sf=\empty
         \ifhmode\edef\#sf{\spacefactor=\the\spacefactor}\/\fi
         #1\@sf\vfootnote{#1}}
\gdef\vfootnote#1{\insert\footins\bgroup
         \ifnum\dimnota=1  \eightpoint\fi
         \ifnum\dimnota=2  \ninepoint\fi
         \ifnum\dimnota=0  \tenpoint\fi
         \interlinepenalty=\interfootnotelinepenalty
         \splittopskip=\ht\strutbox
         \splitmaxdepth=\dp\strutbox \floatingpenalty=20000
         \leftskip=\oldssposta \rightskip=\olddsposta
         \spaceskip=0pt \xspaceskip=0pt
         \ifnum\sinnota=0   \textindent{#1}\fi
         \ifnum\sinnota=1   \item{#1}\fi
         \footstrut\futurelet\next\fo@t}
\gdef\fo@t{\ifcat\bgroup\noexpand\next \let\next\f@@t
             \else\let\next\f@t\fi \next}
\gdef\f@@t{\bgroup\aftergroup\@foot\let\next}
\gdef\f@t#1{#1\@foot} \gdef\@foot{\strut\egroup}
\gdef\footstrut{\vbox to\splittopskip{}}
\skip\footins=\bigskipamount
\count\footins=1000  \dimen\footins=8in
\catcode`@=12
\tenpoint
\ifnum\unoduecol=1 \hsize=\tothsize   \fullhsize=\tothsize \fi
\ifnum\unoduecol=2 \hsize=\collhsize  \fullhsize=\tothsize \fi
\global\let\lrcol=L      \ifnum\unoduecol=1
\output{\plainoutput{\ifnum\tipbnota=2 \clearnmbnota\fi}} \fi
\ifnum\unoduecol=2 \output{\if L\lrcol
     \global\setbox\leftcolumn=\columnbox
     \global\setbox\fotlinebb=\line{\hfill\the\footline\hfill}
     \global\setbox\hedlinebb=\line{\hfill\the\headline\hfill}
     \advancepageno  \global\let\lrcol=R
     \else  \doubleformat \global\let\lrcol=L \fi
     \ifnum\outputpenalty>-20000 \else\dosupereject\fi
     \ifnum\tipbnota=2\clearnmbnota\fi }\fi
\def\ifdoublepage{\ifnum\unoduecol=2 }
\gdef\yespagenumbers{\footline={\hss\tenrm\folio\hss}}
\gdef\ciao{ \ifnum\fdefcontre=1 \endfdef\fi
     \par\vfill\supereject \ifnum\unoduecol=2
     \if R\lrcol  \headline={}\nopagenumbers\null\vfill\eject
     \fi\fi \end}

\newskip\olddsposta \newskip\oldssposta
\global\oldssposta=\leftskip \global\olddsposta=\rightskip

\def\filldots{\leaders\hbox to 1em{\hss.\hss}\hfill}
\def\inquadrb#1 {\vbox {\hrule  \hbox{\vrule \vbox {\vskip .2cm
    \hbox {\ #1\ } \vskip .2cm } \vrule  }  \hrule} }
 \def\newline{\hfil\break}
\def\jump{\vskip\baselineskip} \newskip\iinnffrr
\def\sjump{\iinnffrr=\baselineskip
          \divide\iinnffrr by 2 \vskip\iinnffrr}
\def\bjump{\vskip\baselineskip \vskip\baselineskip}
\newcount\nmbnota  \def\clearnmbnota{\global\nmbnota=0}
\newcount\tipbnota \def\letterfootnote{\global\tipbnota=1}

\def\note#1{\global\advance\nmbnota by 1 \ifnum\tipbnota=1
    \footnote{$^{\rm\nttlett}$}{#1} \else {\ifnum\tipbnota=2
    \footnote{$^{\nttsymb}$}{#1}
    \else\footnote{$^{\the\nmbnota}$}{#1}\fi}\fi}
\def\nttlett{\ifcase\nmbnota \or a\or b\or c\or d\or e\or f\or
g\or h\or i\or j\or k\or l\or m\or n\or o\or p\or q\or r\or
s\or t\or u\or v\or w\or y\or x\or z\fi}
\def\nttsymb{\ifcase\nmbnota \or\dag\or\sharp\or\ddag\or\star\or
\natural\or\flat\or\clubsuit\or\diamondsuit\or\heartsuit
\or\spadesuit\fi}   \clearnmbnota
\def\numberfootnote{\global\tipbnota=0} \numberfootnote
\def\setnote#1{\expandafter\xdef\csname#1\endcsname{
\ifnum\tipbnota=1 {\rm\nttlett} \else {\ifnum\tipbnota=2
{\nttsymb} \else \the\nmbnota\fi}\fi} }
\newcount\nbmfig  \def\clearnbmfig{\global\nbmfig=0}
\gdef\figure{\global\advance\nbmfig by 1
      {\rm fig. \the\nbmfig}}   \clearnbmfig
\def\setfig#1{\expandafter\xdef\csname#1\endcsname{fig. \the\nbmfig}}
 \def\endformula{\eqno\numero $$}
 \def\efr{\endformula}
\newcount\frmcount \def\clearfrmcount{\global\frmcount=0}
\def\numero{\global\advance\frmcount by 1   \ifnum\indappcount=0
  {\ifnum\cpcount <1 {\hbox{\rm (\the\frmcount )}}  \else
  {\hbox{\rm (\the\cpcount .\the\frmcount )}} \fi}  \else
  {\hbox{\rm (\applett .\the\frmcount )}} \fi}
\def\nameformula#1{\global\advance\frmcount by 1%
\ifnum\draftnum=0  {\ifnum\indappcount=0%
{\ifnum\cpcount<1\xdef\spzzttrra{(\the\frmcount )}%
\else\xdef\spzzttrra{(\the\cpcount .\the\frmcount )}\fi}%
\else\xdef\spzzttrra{(\applett .\the\frmcount )}\fi}%
\else\xdef\spzzttrra{(#1)}\fi%
\expandafter\xdef\csname#1\endcsname{\spzzttrra}
\eqno \hbox{\rm\spzzttrra} $$}
\def\nfr{\nameformula}    
\def\nameali#1{\global\advance\frmcount by 1%
\ifnum\draftnum=0  {\ifnum\indappcount=0%
{\ifnum\cpcount<1\xdef\spzzttrra{(\the\frmcount )}%
\else\xdef\spzzttrra{(\the\cpcount .\the\frmcount )}\fi}%
\else\xdef\spzzttrra{(\applett .\the\frmcount )}\fi}%
\else\xdef\spzzttrra{(#1)}\fi%
\expandafter\xdef\csname#1\endcsname{\spzzttrra}
  \hbox{\rm\spzzttrra} }      \clearfrmcount
\newcount\cpcount \def\clearcpcount{\global\cpcount=0}
\newcount\subcpcount \def\clearsubcpcount{\global\subcpcount=0}
\newcount\appcount \def\clearappcount{\global\appcount=0}
\newcount\indappcount \def\clearindappcount{\indappcount=0}
\newcount\sottoparcount 

\def\applett{\ifcase\appcount  \or {A}\or {B}\or {C}\or
{D}\or {E}\or {F}\or {G}\or {H}\or {I}\or {J}\or {K}\or {L}\or
{M}\or {N}\or {O}\or {P}\or {Q}\or {R}\or {S}\or {T}\or {U}\or
{V}\or {W}\or {X}\or {Y}\or {Z}\fi    \ifnum\appcount<0
\immediate\write16 {Panda ERROR - Appendix: counter "appcount"
out of range}\fi  \ifnum\appcount>26  \immediate\write16 {Panda
ERROR - Appendix: counter "appcount" out of range}\fi}
\clearappcount  \clearindappcount \newcount\connttrre
\def\clearconnttrre{\global\connttrre=0} \newcount\countref
\def\clearcountref{\global\countref=0} \clearcountref
\def\chapter#1{\global\advance\cpcount by 1 \clearfrmcount
                 \goodbreak\null\vbox{\jump\nobreak
                 \clearsubcpcount\clearindappcount
                 \itemitem{\ttaarr\the\cpcount .\qquad}{\ttaarr #1}
                 \par\nobreak\jump\sjump}\nobreak}
\def\section#1{\global\advance\subcpcount by 1 \goodbreak\null
               \vbox{\sjump\nobreak\ifnum\indappcount=0
                 {\ifnum\cpcount=0 {\itemitem{\ppaarr
               .\the\subcpcount\quad\enskip\ }{\ppaarr #1}\par} \else
                 {\itemitem{\ppaarr\the\cpcount .\the\subcpcount\quad
                  \enskip\ }{\ppaarr #1} \par}  \fi}
                \else{\itemitem{\ppaarr\applett .\the\subcpcount\quad
                 \enskip\ }{\ppaarr #1}\par}\fi\nobreak\jump}\nobreak}
\clearsubcpcount
\def\appendix#1{\global\advance\appcount by 1 \clearfrmcount
                  \goodbreak\null\vbox{\jump\nobreak
                  \global\advance\indappcount by 1 \clearsubcpcount
          \itemitem{ }{\hskip-40pt\ttaarr Appendix\ #1}
             \nobreak\jump\sjump}\nobreak}
\clearappcount \clearindappcount
\def\references{\goodbreak\null\vbox{\jump\nobreak
   \itemitem{}{\ttaarr References} \nobreak\jump\sjump}\nobreak}

\clearcpcount\clearcountref

\def\setchap#1{\ifnum\indappcount=0{\ifnum\subcpcount=0%
\xdef\spzzttrra{\the\cpcount}%
\else\xdef\spzzttrra{\the\cpcount .\the\subcpcount}\fi}
\else{\ifnum\subcpcount=0 \xdef\spzzttrra{\applett}%
\else\xdef\spzzttrra{\applett .\the\subcpcount}\fi}\fi
\expandafter\xdef\csname#1\endcsname{\spzzttrra}}
\newcount\draftnum \newcount\ppora   \newcount\ppminuti
\global\ppora=\time   \global\ppminuti=\time
\global\divide\ppora by 60  \draftnum=\ppora
\multiply\draftnum by 60    \global\advance\ppminuti by -\draftnum
\def\droggi{\number\day /\number\month /\number\year\ \the\ppora
:\the\ppminuti}     \global\draftnum=0
\def\draftcomment#1{\ifnum\draftnum=0 \relax \else
{\ {\bf ***}\ #1\ {\bf ***}\ }\fi} 
%
%
\catcode`@=11
\gdef\Ref#1{\expandafter\ifx\csname @rrxx@#1\endcsname\relax%
{\global\advance\countref by 1    \ifnum\countref>200
\immediate\write16 {Panda ERROR - Ref: maximum number of references
exceeded}  \expandafter\xdef\csname @rrxx@#1\endcsname{0}\else
\expandafter\xdef\csname @rrxx@#1\endcsname{\the\countref}\fi}\fi
\ifnum\draftnum=0 \csname @rrxx@#1\endcsname \else#1\fi}
\gdef\beginref{\ifnum\draftnum=0  \gdef\Rref{\fairef}
\gdef\endref{\scriviref} \else\relax\fi
\ifx\risposta\mplarisposta \ninepoint \fi
\parskip 2pt plus.2pt \baselineskip=12pt}
\def\Reflab#1{[#1]} \gdef\Rref#1#2{\item{\Reflab{#1}}{#2}}
\gdef\endref{\relax}  \newcount\conttemp
\gdef\fairef#1#2{\expandafter\ifx\csname @rrxx@#1\endcsname\relax
{\global\conttemp=0 \immediate\write16 {Panda ERROR - Ref: reference
[#1] undefined}} \else
{\global\conttemp=\csname @rrxx@#1\endcsname } \fi
\global\advance\conttemp by 50  \global\setbox\conttemp=\hbox{#2} }
\gdef\scriviref{\clearconnttrre\conttemp=50
\loop\ifnum\connttrre<\countref \advance\conttemp by 1
\advance\connttrre by 1
\item{\Reflab{\the\connttrre}}{\unhcopy\conttemp} \repeat}
\clearcountref \clearconnttrre
\catcode`@=12
\ifx\risposta\mplarisposta \def\Reflab#1{#1.} \letterfootnote \fi

\def\slashchar#1{\setbox0=\hbox{$#1$} \dimen0=\wd0
     \setbox1=\hbox{/} \dimen1=\wd1 \ifdim\dimen0>\dimen1
      \rlap{\hbox to \dimen0{\hfil/\hfil}} #1 \else
      \rlap{\hbox to \dimen1{\hfil$#1$\hfil}} / \fi}
\ifx\oldchi\undefined \let\oldchi=\chi
  \def\cchi{{\raise 1pt\hbox{$\oldchi$}}} \let\chi=\cchi \fi
  
\def\del{\partial}   

\def\frac#1#2{{\textstyle{#1 \over #2}}}

\def\half{\ifinner {\scriptstyle {1 \over 2}}\else {1 \over 2} \fi}

\def\simge{\rlap{\raise 2pt \hbox{$>$}}{\lower 2pt \hbox{$\sim$}}}
\def\simle{\rlap{\raise 2pt \hbox{$<$}}{\lower 2pt \hbox{$\sim$}}}

\def\vbig#1#2{{\vbigd@men=#2\divide\vbigd@men by 2%
\hbox{$\left#1\vbox to \vbigd@men{}\right.\n@space$}}}

%
%
\newcount\fdefcontre \newcount\fdefcount \newcount\indcount
\newread\filefdef  \newread\fileftmp  \newwrite\filefdef
\newwrite\fileftmp     \def\strip#1*.A {#1}
\def\futuredef#1{\beginfdef
\expandafter\ifx\csname#1\endcsname\relax%
{\immediate\write\fileftmp {#1*.A}
\immediate\write16 {Panda Warning - fdef: macro "#1" on page
\the\pageno \space undefined}
\ifnum\draftnum=0 \expandafter\xdef\csname#1\endcsname{(?)}
\else \expandafter\xdef\csname#1\endcsname{(#1)} \fi
\global\advance\fdefcount by 1}\fi   \csname#1\endcsname}

\def\beginfdef{\ifnum\fdefcontre=0
\immediate\openin\filefdef \jobname.fdef
\immediate\openout\fileftmp \jobname.ftmp
\global\fdefcontre=1  \ifeof\filefdef \immediate\write16 {Panda
WARNING - fdef: file \jobname.fdef not found, run TeX again}
\else \immediate\read\filefdef to\spzzttrra
\global\advance\fdefcount by \spzzttrra
\indcount=0      \loop\ifnum\indcount<\fdefcount
\advance\indcount by 1   \immediate\read\filefdef to\spezttrra
\immediate\read\filefdef to\sppzttrra
\edef\spzzttrra{\expandafter\strip\spezttrra}
\immediate\write\fileftmp {\spzzttrra *.A}
\expandafter\xdef\csname\spzzttrra\endcsname{\sppzttrra}
\repeat \fi \immediate\closein\filefdef \fi}
\def\endfdef{\immediate\closeout\fileftmp   \ifnum\fdefcount>0
\immediate\openin\fileftmp \jobname.ftmp
\immediate\openout\filefdef \jobname.fdef
\immediate\write\filefdef {\the\fdefcount}   \indcount=0
\loop\ifnum\indcount<\fdefcount    \advance\indcount by 1
\immediate\read\fileftmp to\spezttrra
\edef\spzzttrra{\expandafter\strip\spezttrra}
\immediate\write\filefdef{\spzzttrra *.A}
\edef\spezttrra{\string{\csname\spzzttrra\endcsname\string}}
\iwritel\filefdef{\spezttrra}
\repeat  \immediate\closein\fileftmp \immediate\closeout\filefdef
\immediate\write16 {Panda Warning - fdef: Label(s) may have changed,
re-run TeX to get them right}\fi}
\def\iwritel#1#2{\newlinechar=-1
{\newlinechar=`\ \immediate\write#1{#2}}\newlinechar=-1}
\global\fdefcontre=0 \global\fdefcount=0 \global\indcount=0
%
%
\null
%
%
%
%

%
\loadamsmath
%
\def\CP{{\Bbb C}{\rm P}^n }
\pageno=0\baselineskip=14pt
\nopagenumbers{
\line{\hfill CERN-TH.7425/94}
\line{\hfill SWAT/93-94/39}
\line{\hfill\tt hep-th/9409141}
\line{\hfill September 1994}
\ifdoublepage \bjump\bjump\bjump\bjump\else\vfill\fi
\centerline{\capsone The exact mass-gap of the supersymmetric o($N$)
sigma model}
\bjump\bjump
\centerline{\scaps Jonathan M. Evans\footnote{$^{*}$}{Supported by a
fellowship from the EU Human Capital and Mobility programme.}
Timothy J.~Hollowood\footnote{$^{**}$}{On leave from:
Department of Physics, University of Wales, Swansea, SA2
8PP, U.K.}}
\sjump
\sjump
\centerline{\sl CERN-TH, CH-1211 Geneva 23, Switzerland.}
\centerline{\tt evansjm@surya11.cern.ch, hollow@surya11.cern.ch}
\bjump\bjump\bjump
\ifdoublepage
\vfill
\noindent
\line{CERN-TH.7425/94\hfill}
\line{September 1994\hfill}
\eject\null\vfill\fi
\centerline{\capsone ABSTRACT}\sjump
A formula for the mass-gap of the supersymmetric O($N$) sigma
model ($N>4$) in two dimensions is derived:
$m/\Lambda_{\overline{\rm MS}}=2^{2\Delta}\sin(\pi\Delta)/(\pi\Delta)$,
where $\Delta=1/(N-2)$
and $m$ is the mass of the fundamental vector particle in the theory.
This result is obtained by comparing two expressions for the
free-energy density in
the presence of a coupling to a conserved charge; one expression is computed
from the exact S-matrix of Shankar and Witten via the the thermodynamic Bethe
ansatz and the other is computed using conventional perturbation theory.
These calculations provide a stringent test of the S-matrix, showing that it
correctly reproduces the universal part of the beta-function
and resolving the problem of CDD ambiguities.
\sjump\vfill
\ifdoublepage \else
\noindent
\line{CERN-TH.7425/94\hfill}
\line{September 1994\hfill}\fi
\eject}
\yespagenumbers\pageno=1
%
%
\def\t{\theta}

\chapter{Introduction}

There are many two-dimensional field theories which are quantum integrable
and hence, following conventional wisdom, which are thought to be described
by an exact factorizable S-matrix.
Of particular interest are the theories which generate their mass dynamically,
like the O($N$) sigma model, since they share many of the features of QCD in
four dimensions.
In general the exact S-matrices for these models are---if truth be
told---conjectures which are postulated on the basis of symmetries and
various physical properties which are encoded as axioms of S-matrix theory.
The property of factorization is then
enough in many cases to determine the S-matrix up to ambiguities of
CDD type [\Ref{ZZ}].
It is important to scrutinize these S-matrices and find
ways of checking whether they do indeed describe the field theories
for which they are designed.

Given a factorizable S-matrix, it is possible to find a set of
integral equations, called the Thermodynamic Bethe Ansatz (TBA)
equations, which determine the free-energy of the theory on a cylinder in
the presence of a chemical potential which couples to a conserved
charge in the model. It was realized some time ago,
in the context of the SU($N$) principal chiral models,
that it is possible to use the TBA equations
to extract the universal coefficients of the beta-function
directly from the S-matrix [\Ref{PW},\Ref{W}].
The idea is to compare the free-energy extracted from the TBA
equations at zero temperature (hence on the plane) to the same
quantity evaluated in the asymptotic regime where the chemical potential is
large and so perturbation theory is
valid. Comparing these expressions provides a stringent test of the
proposed S-matrix and also yields an exact value for the mass-gap of
the theory, by which we mean the ratio of some chosen physical mass $m$ to the
$\Lambda$-parameter of perturbation theory.

This strategy has been applied to a series of models:
[\Ref{HMN},\Ref{HN}] for the O($N$) sigma
model; [\Ref{BNNW}] for the SU($N$) principal chiral model;
[\Ref{THIII}] for the SO($N$) and Sp($N$) principal chiral models;
[\Ref{CGN}] for the SU($N$) chiral Gross-Neveu models;
[\Ref{FNW}] for the O($N$) Gross-Neveu models; and [\Ref{EH}] for
integrable sigma models on an SU(2) group manifold with torsion.
In each case the exact mass-gap was extracted and the S-matrix tested.
It is perhaps significant that in each case the minimal S-matrix (the
S-matrix with the minimum number of poles and zeros on the physical
strip consistent with physical requirements) was found to be consistent.
It is clearly very useful to have such exact results for mass gaps
because they provide a
remarkable opportunity to test the efficacy of lattice simulations or other
non-perturbative approaches.
In addition to this, however, the results of [\Ref{HMN}-\Ref{EH}]
provide valuable concrete illustrations
of the correctness of conventional beliefs regarding the character of
asymptotically-free theories with dynamical mass generation.

In this paper we shall apply the techniques described above to the
supersymmetric O($N$) sigma model [\Ref{Wit},\Ref{SW},\Ref{DF}].
The application of these methods to a supersymmetric integrable field
theory raises some novel issues, as we shall see. In technical terms we must
face a diagonalization problem for the TBA equations which does not arise
in the purely bosonic case.
In many respects the O($N$) theories which we consider here are the simplest
family of supersymmetric sigma models.
There is another very well-known family of super sigma models based on $\CP$,
which in fact exhibit extended ($N=2$) supersymmetry, and which have a
richer structure than the O($N$) models at both the classical
and quantum levels. These $\CP$ models give rise to a more complicated
set of TBA equations and they will be treated
in a sequel to this paper [\Ref{EH2}].

\chapter{The model and its S-matrix}

The lagrangian density of the supersymmetric O($N$) model is [\Ref{Wit}]
$$
{\cal L}={1\over2g}\left[(\partial_\mu
n_a)^2+i\bar\psi_a\slashchar\partial\psi_a+\frac{1}{4}\left(
\bar\psi_a\psi_a\right)^2\right],
\nfr{lag}
where $n_a$ and $\psi_a$ are an $N$-component real scalar field and
an $N$-component Majorana fermion respectively satisfying the constraints
$n\cdot n=1$ and $n\cdot\psi=0$.
We work throughout in two-dimensional Minkowski space and our
conventions agree with those of [\Ref{Wit},\Ref{REN}].
We shall consider only the cases $N>4$ (the O(3) model fits more
naturally into the family of $\CP$ theories discussed in [\Ref{EH2}]
and the O(4) model is in fact equivalent to the principal chiral model
based on SU(2)).
The theory \lag\ has a global O($N$)
symmetry and a global $N=1$ supersymmetry. Notice that the bosonic
part of the theory is just the O($N$) sigma model, the fermionic
part is the O($N$) Gross-Neveu model, and the coupling between the bosons and
fermions is due solely to the constraint.

The two-loop beta-function for this model and the corresponding behaviour
of the running coupling constant can be written
$$\eqalign{
\beta(g) & = - \beta_1 g^2 - \beta_2 g^3 + {\cal O} (g^4),\cr
&{\rm so}\quad{1 \over g(\mu / \Lambda)} = \beta_1 \ln {\mu \over \Lambda}
+ {\beta_2 \over \beta_1} \ln \ln {\mu\over \Lambda}
+ {\cal O} \left( {\ln \ln (\mu /\Lambda) \over \ln (\mu / \Lambda)} \right ),
\cr
& {\rm where} \quad \beta_1 = (N-2) / 2\pi , \quad \beta_2 = 0.\cr
}
\nfr{RG}
Note that the first coefficient of the beta-function
coincides with the result for the purely bosonic O($N$) model, whereas
the second coefficient vanishes,
unlike the purely bosonic or Gross-Neveu cases.
The values of both these coefficients can be deduced from general results
[\Ref{REN}] concerning supersymmetric sigma-models on locally symmetric
spaces (see also [\Ref{GVZ}] and references cited there for
details of subsequent work). We see from the beta-function that the theory is
asymptotically free with dynamical mass generation.

The integrability of the supersymmetric O($N$) theory was studied over 16
years ago by Shanker and Witten [\Ref{SW}] who, following [\Ref{ZZ}],
proposed a factorizable S-matrix to describe the scattering of the
fundamental multiplet of particles in the model.
It is expected that this fundamental multiplet will appear as a
massive supersymmetric doublet which transforms as a vector under the O($N$)
symmetry. We will denote the corresponding quantum states by
$|a,i,\t\rangle$, where $i=0,1$ labels a boson, fermion respectively,
$a$ is the O($N$) vector index and $\t$ is the rapidity of
the particle, so that its velocity is $v={\rm tanh}(\t)$.
The full spectrum of the theory will also contain bound states
of this fundamental multiplet, but detailed knowledge of these will not be
important for our purposes.

The $S$-matrix conjectured by Shankar and Witten
has a very particular form in which the
supersymmetric and O($N$) degrees of freedom are factored.
This means that the
two-body $S$-matrix elements, from which all others follow,
can be written [\Ref{SHOU}]
$$
\langle c,k,\t_2;d,l,\t_1,{\rm out}|a,i,\t_1;b,j,\t_2,{\rm in}\rangle=
S_{\rm SUSY}(\t_1-\t_2)_{ij}^{kl}S_{\rm GN}(\t_1-\t_2)_{ab}^{cd}.
\nfr{SM}
Here the O($N$) part is the factorizable S-matrix of the
fundamental vector particle of the O($N$) Gross-Neveu model
[\Ref{ZZ},\Ref{SMGN}]:
$$
S_{\rm GN}(\t)_{ab}^{cd}=Y_1(\t)\left[\delta^{ad}\delta^{bc}-{2\pi
i\Delta\over i\pi-\t}\delta^{ab}\delta^{cd}-{2\pi
i\Delta\over\t}\delta^{ac}\delta^{bd}\right],
\efr
where $\Delta=1/(N-2)$ and the unitarizing-crossing scalar factor is
$Y_1(\t)=R_1(\t)R_1(i\pi-\t)$ with
$$
R_1(\t)={\Gamma(-\Delta-i\t/2\pi)\Gamma(\half-i\t/2\pi)\over\Gamma(-i\t
/2\pi)\Gamma(\half-\Delta-i\t/2\pi)}.
\efr
The supersymmetric part of the S-matrix has the form
$$
S_{\rm SUSY}(\t)=Y_2(\t)
\pmatrix{1+2i{\sin(\pi\Delta)\over\sinh\t}&0&0&i{\sin(\pi\Delta)\over
\cosh(\t/2)}\cr
0&i{\sin(\pi\Delta)\over\sinh(\t/2)}&1&0\cr
0&1&i{\sin(\pi\Delta)\over\sinh(\t/2)}&0\cr
-i{\sin(\pi\Delta)\over\cosh(\t/2)}&0&0&-1+2i{\sin(\pi\Delta)\over\sinh\t}
\cr},
\nfr{SUSYSM}
in which the rows and columns are labelled in the order
$(0,0),(0,1),(1,0),(1,1)$.
In this case the scalar factor is $Y_2(\t)=R_2(\t)R_2(i\pi-\t)$
where\note{When comparing with the
expressions in [\Ref{SW}] it is helpful to notice that we are
including the simple pole term in the O($N$) rather than in the
supersymmetric factor; the opposite choice was made in [\Ref{SW}] but
the net results are obviously equivalent.}
$$\eqalign{
&R_2(\t)={\Gamma(-i\t/2\pi)\over\Gamma(\half-i\t/2\pi)}\cr
&\times\prod_{j=1}^\infty{\Gamma(-\Delta-i\t/2\pi+j)\Gamma(-i\t/2\pi+
\Delta+j-1)
\Gamma^2(-i\t/2\pi+j-\half)\over\Gamma(-\Delta-i\t/2\pi+j+\half)
\Gamma(-i\t/2\pi+\Delta+j-\half)\Gamma^2(-i\t/2\pi+j-1)}.\cr}
\efr
It is important that the ordering of the particles in the final
state is taken so that the particle of rapidity $\t_2$ is
to the left of the particle with rapidity $\t_1$;
it is only this ``modified'' S-matrix that displays the factorization of
supersymmetric and bosonic degrees of freedom
as described in [\Ref{SHOU}].

The S-matrix \SM\ is a ``minimal'' expression in the sense that it has
the minimum number of poles and zeros on the physical strip (the
region $0\leq{\rm Im}(\t)\leq\pi$) consistent with the
requirements of symmetry, the axioms of S-matrix theory, and the
implementation of the bootstrap procedure.
It is well-known, however, that solutions to these conditions
are ambiguous precisely up to so-called CDD factors,
which for the present model take the form
$$
{\sinh\t-i\sin(\pi\Delta(2-\alpha))
\over
\sinh\t+i\sin(\pi\Delta(2-\alpha))}\cdot
{\sinh\t-i\sin(\pi\Delta\alpha)
\over
\sinh\t+i\sin(\pi\Delta\alpha)},
\nfr{CDDF}
where $0<\alpha<2$ is a constant.
Multiplying \SM\ by any number of factors of this type does not
introduce any additional poles on the physical strip and
respects the internal consistency of the S-matrix.
An important aspect of the results we shall obtain is that they
will resolve this possible ambiguity in favour of the minimal choice \SM .
This is established by simply adopting the minimal S-matrix \SM\
and checking that the results derived from it agree exactly
with perturbation theory, whereas any additional CDD factors would
alter substantially the result of our calculation.

For completeness we mention how the entire spectrum of the model can
be determined.
The minimal S-matrix of the fundamental particles
has a simple pole on the physical strip at $\t=2\pi
i\Delta$ which corresponds to a bound state transforming in a
reducible representation which is the sum of the antisymmetric tensor
and singlet representations of O($N$). Continuing the
bootstrap in this way one finds a spectrum of bound-states
which is identical to the O($N$) Gross-Neveu model, namely $m_r=m\sin(\pi
r\Delta)/\sin(\pi\Delta)$, $1\leq r<(N/2)-1$, apart from the fact that here
each particle carries additional supersymmetric quantum numbers. We shall only
require the S-matrix elements of the fundamental particle for our
calculation.

\chapter{Coupling to a conserved charge}

To follow the logic of [\Ref{HMN}--\Ref{EH}] one couples the theory
to a background field $h$ via a conserved charge $Q$ corresponding to some
generator of a global symmetry in the model. The field $h$ acts as a
chemical potential for eigenstates of $Q$, and the idea is
to compute the corresponding free-energy per unit volume $f(h)$ in the
ultra-violet, large $h$, regime.
This is clearly equivalent to calculating the ground state energy
density of the system
with the Hamiltonian modified from $H$ to $H-hQ$.
In fact we are interested only in the finite difference
$\delta f (h) = f(h) - f(0)$. The TBA equations which follow from the
S-matrix in principle determine this quantity exactly,
however one cannot usually solve them for all $h$.
If the system is sufficiently simple, one can obtain an expansion
valid in the asymptotic regime $h \gg m $, yielding a result of the form
$\delta f (h) = h^2 F_1 (h / m)$.
But $h \gg m$ is also exactly the regime in which conventional
perturbation theory can be applied and such a calculation yields an expression
of the form $\delta f (h) = h^2 F_2 (h / \Lambda)$.
By equating $F_1 (h/m) = F_2 (h/\Lambda)$ one obtains a powerful
check of the consistency
of the proposed S-matrix with perturbation theory,
and one extracts the mass-gap $m/\Lambda$.

A crucial part of this procedure is the precise choice of $Q$.
The strategy followed in [\Ref{HMN}--\Ref{EH}]
is to choose $Q$ so that it has a unique largest
eigenvalue, $+1$ say, corresponding to some unique
fundamental particle state. Then one argues that for large $h$ only this
particle state appears in the ground state of the new Hamiltonian
$H - hQ$ and the TBA system is thereby reduced to a single integral equation.
In fact this argument assumes that bound states which may have the same
$Q$ eigenvalue will not contribute either, on the grounds that they will have
a smaller charge/mass ratio. In principle this should follow from a rigorous
analysis of the full TBA system, but in practice the complexity of the full
system means that the assumption must usually be taken as a working
hypothesis which is ultimately vindicated by the final results
[\Ref{HMN}--\Ref{EH}].

A novel feature of dealing with a supersymmetric system is that
the best one can do is to choose $Q$ so that its largest eigenvalue picks
out a supermultiplet of
states rather than a single state (since $Q$ commutes with supersymmetry).
In the present case we can choose, for example, a charge with
$$
Q_{12}=i,\qquad Q_{21}=-i,
\nfr{CH}
and all other components zero. It is then exactly the doublet of states
$(|1,j,\t\rangle+i|2,j,\t\rangle)/\sqrt2$ which have eigenvalue +1 under $Q$
(whereas all the other eigenstates have eigenvalues 0 or $-1$). Our
hypothesis is that only these particles will appear in the ground-state
of the new Hamiltonian $H - hQ$.
The scattering of these states amongst themselves is
elastic in the space of O($N$) quantum numbers but it is
still non-diagonal in the supersymmetric subspace. The
explicit S-matrix for these states is
$$
S(\t)_{ij}^{kl}=U(\t)S_{\rm SUSY}(\t)_{ij}^{kl},
\efr
where the scalar factor comes from the elastic scattering of the
O($N$) part of the S-matrix and has the form
$$
U(\t)={\Gamma(1+i\t/2\pi)\Gamma(\half-i\t/2\pi)\Gamma(1-\Delta-i\t/2\pi)
\Gamma(\half-\Delta+i\t/2\pi)\over
\Gamma(1-i\t/2\pi)\Gamma(\half+i\t/2\pi)\Gamma(1-\Delta+i\t/2\pi)
\Gamma(\half-\Delta-i\t/2\pi)}.
\nfr{DEFU}
As a result of the non-trivial scattering amongst the supersymmetric degrees
of freedom we shall have to confront a set of two coupled TBA equations
instead of a single integral equation as in
[\Ref{HMN}--\Ref{EH}].

Another important aspect of the choice of $Q$ is that it can drastically
affect the
nature of the expressions $F_1(h/m)$ and $F_2(h/\Lambda)$ which we are trying
to calculate.
Experience with other models [\Ref{HMN}--\Ref{FNW}] suggests that
bosonic theories generally require perturbation theory to just one loop to
extract the mass-gap
(although the theories in [\Ref{EH}] are an
exception) whereas purely fermionic theories seem to require three
loop calculations.
Fortunately, we shall find that a one-loop
calculation suffices for the choice of $Q$ given above.

\chapter{Free-energy from perturbation theory}

The coupling to the charge $Q$ by means of the replacement
$H \rightarrow H - hQ$ can be achieved at the lagrangian level by
taking ${\cal L}$ and making the
replacement $\del_0 \rightarrow \del_0 + ih Q$ which resembles
a covariant derivative.
Having introduced this coupling, we wish to compute
the free-energy to one-loop, which means that it is sufficient to
expand the lagrangian to quadratic order in some set of independent fields.
Since we are interested only in the difference
$\delta f(h)=f(h)-f(0)$ we can also ignore any fields which
do not couple to $h$ at this order.
If we solve the bosonic constraint
by writing $(n_1 , n_2) = (\cos \theta, \sin \theta) \sqrt{ 1 - \pi^2 }$
where $\pi=(n_3,n_4,\ldots,n_N)$ then it is easy to see that
the field $\theta$ decouples, as do all the fermionic degrees of freedom,
and it suffices to consider the lagrangian
$$
{\cal L}_{\rm 1-loop}=
{1\over2g}\left(\partial_\mu\pi\right)^2+ {h^2\over2
g}\left(1-\pi^2\right).
\efr
This leaves exactly the same calculation as encountered in
the bosonic O($N$) sigma model [\Ref{HN}].

Using dimensional regularization with the
$\overline{\rm MS}$-scheme one finds the result
$$
\delta f (h) =
-{h^2 \over 2g} + {h^2 \over 8 \pi} (N-2) (1 - \ln (h^2 / \mu^2) )
+ {\cal O} (g)
\nfr{CCE}
in terms of the running coupling $g(\mu / \Lambda_{ \overline{ {\rm MS} } } )$.
We can use the fact that $\delta f (h)$ is a physical quantity, and therefore
RG-invariant, to extract $\beta_1$ from the expression above and to check that
it agrees with \RG . (One cannot extract $\beta_2$ from this result alone
because it is only valid to one-loop.)
Now to compare with the TBA result we must substitute the explicit
expression for the running coupling to two loops given in \RG.
For future reference we first write
the result in a way which reveals the functional dependence on the
coefficients of the beta-function:
$$
\delta f(h)=-h^2{\beta_1\over2}\left[\ln{h\over
\Lambda_{\overline{\rm MS}}}-{1\over2}+{\beta_2\over\beta_1^2}
\ln\ln{h
\over\Lambda_{\overline{\rm MS}}}+{\cal O}\left
({\ln\ln(h/\Lambda_{\overline{\rm
MS}})\over\ln(h/\Lambda_{\overline{\rm MS}})}\right)\right].
\efr
On taking the specific values of these coefficients given in \RG\ we
obtain
$$
\delta f(h)=-(N-2){h^2\over4\pi}\left[\ln{h\over
\Lambda_{\overline{\rm MS}}}-{1\over2}+{\cal O}\left
({\ln\ln(h/\Lambda_{\overline{\rm
MS}})\over\ln(h/\Lambda_{\overline{\rm MS}})}\right)\right],
\nfr{FEP}
which can be contrasted with the result for the bosonic O($N$) sigma model
(equation (18) of [\Ref{HN}]). Notice that the absense of the
$\ln\ln(h/\Lambda_{\overline{\rm MS}})$ term is due to the vanishing of
the second coefficient of the beta-function. We also remark that the
number of terms in the expansion of $\delta f(h)$ in \FEP\ will
suffice to exact the mass-gap; this is directly related to the
existence of the ``tree-level'' ${\cal O}(1/g)$ term in \CCE. In
contrast to this, for a fermionic model there is no tree-level
contribution and a three-loop calculation is needed to extract the mass-gap.

\chapter{Free-energy from the S-matrix}

We now write down the TBA equations for the model and find
their solution in the limit $h\gg m$. Recall our
hypothesis that with the coupling to the charge \CH\ only the states
$(|1,j,\t\rangle+i|2,j,\t\rangle)/\sqrt2$ will contribute to the
ground-state; this allows us to avoid the difficult problem of solving
the full TBA equations including the O($N$) magnon system.
But these states are, after all, a supersymmetric multiplet
and the scattering is non-diagonal in this subspace. Fortunately, the
diagonalization of the relevant system of equations has been performed
recently by Ahn [\Ref{AHN}] who
exploited the equivalence of the problem to that of diagonalizing the
transfer matrix of the eight vertex model at the free fermion point.

The set of equations derived in [\Ref{AHN}] relates the
density of single particle states in rapidity space $\varrho(\t)$ to the
density of occupied single particle states $\sigma(\t)$.
The equations involve
additional densities $P_+(\t)$ and $P_-(\t)$ corresponding to a
``supersymmetric magnon'':
$$\eqalign{
&\varrho(\t)={m\over2\pi}\cosh\t+\Psi*\sigma(\t)+{1\over2}\Phi*[P_+(\t)
-P_-(\t)],\cr &P_+(\t)+P_-(\t)=\Phi*\sigma(\t),\cr}
\nfr{BA}
where $f*g(\t)=\int_{-\infty}^\infty
d\t'f(\t-\t')g(\t')$. The kernels appearing in \BA\ are
$$
\eqalign{
\Psi(\t)&={1\over2\pi}{\rm Im}{d\over d\t}
\ln\left[{U(\t)Y_2(\t)\over\sinh\t}\right],\cr
\Phi(\t)&={1\over2}\cdot{\sin(2\pi\Delta)\over\cosh^2\t-\cos^2
(\pi\Delta)}.\cr}
\efr
We can rewrite \BA\ in a more suggestive form by eliminating
$P_-(\t)$ from the first equation:
$$
{m\over2\pi}\cosh\t=\varrho(\t)-\Phi_{\rm GN}*\sigma(\t)-\Phi*P_+(\t),
\efr
where one finds
$$
\Phi_{\rm GN}(\t)={1\over2\pi}{\rm Im}{d\over d\t}\ln U(\t).
\efr
Notice that in this form the kernel which multiplies $\sigma(\t)$ involves
a contribution only from the Gross-Neveu part of the S-matrix due to
the remarkable cancellation:
$$
{1\over2\pi}{\rm Im}{d\over
d\t}\ln\left[{Y_2(\t)\over\sinh\t}\right]-{1\over2}\Phi*\Phi(\t)=0.
\efr

The passage to the TBA equations proceeds in the usual manner
[\Ref{TBA}]. At finite
temperature $T$ we define the excitation energies of the particle
$\epsilon(\t)$ and the magnon $\xi(\t)$ via
$$
{\sigma(\t)\over\varrho(\t)}={1\over e^{\epsilon(\t)/T}+1},\qquad
{P_-(\t)\over P_+(\t)}=e^{\xi(\t)/T}.
\efr
We shall only require the TBA equations at zero temperature with a chemical
potential $h$ coupled to the particles. In this case
$\epsilon(\pm\t_{\rm F})=0$, where
$\t_{\rm F}$ is the Fermi rapidity, and $\epsilon(\t)$ is negative for
$-\t_{\rm F}<\t<\t_{\rm F}$. The free-energy per unit volume at $T=0$
is given by
$$
\delta f(h)={m\over2\pi}\int_{-\t_{\rm F}}^{\t_{\rm
F}}d\t\,\epsilon(\t)
\cosh\t,
\nfr{FE}
where $\epsilon(\t)$ is the solution of the $T=0$ TBA equations:
$$
\eqalign{
\epsilon(\t)-\Phi_{\rm GN}*\epsilon^-(\t)-\Phi*\xi^-(\t)&=m\cosh\t-h,\cr
\xi(\t)-\Phi*\epsilon^-(\t)&=0,\cr}
\nfr{TBA}
and we have used the notation
$$
f^\pm(\t)=\cases{f(\t)\qquad&$f(\t){>\atop<}0$\cr 0&otherwise.\cr}
\efr
Notice that if we remove the term involving the magnon from the first
TBA equation then it reduces to that encountered in O($N$) Gross-Neveu
model [\Ref{FNW}].

We must now solve \TBA\ and we will implicitly assume that the
solution is unique. The crucial observation is that $\Phi(\t)$ is a
positive kernel; hence the solution has $\xi^+(\t)=0$ and
$\xi^-(\t)=\Phi*\epsilon^-(\t)$ so that the two-dimensional system
reduces to a single equation for $\epsilon(\t)$:
$$
\epsilon^+(\t)+R*\epsilon^-(\t)=m\cosh\t-h.
\nfr{RTBA}
where the kernel is
$$
R(\t)=\delta(\t)-\Phi_{\rm GN}(\t)-\Phi*\Phi(\t).
\efr
This equation differs from that encountered in the O($N$) Gross-Neveu
model by the presence of the term involving $\Phi*\Phi(\t)$ and we shall find
that this drastically alters the nature of the solution.

To determine the behaviour of the solution in the regime $h \gg m$ we
calculate the Fourier transform of the kernel in \RTBA\ and find
$$\eqalign{
&R(\t)=\cr
&\int_{0}^\infty{d\omega\over\pi}\cos(\omega\t)
\left[{\cosh((1-2\Delta)\pi\omega/2)
\over\cosh(\pi\omega/2)}e^{\Delta\pi\omega/2}-{\cosh^2((1-2\Delta)
\pi\omega/2)\over\cosh^2(\pi\omega/2)}\right]\cr
&=\int_0^\infty{d\omega\over\pi}\cos(\omega\t){\cosh((1-2\Delta)\pi
\omega/2)\sinh(\pi\Delta\omega)\over\cosh^2(\pi\omega/2)}e^{\pi\omega/2}.\cr}
\efr
The nature of the solution depends upon whether or not this
Fourier transform vanishes at the origin
[\Ref{FNW}]. We see here that the part that comes from the Gross-Neveu model
does not vanish, but the Fourier transform of the
full kernel, with the effect of supersymmetry
included, does vanish at the origin. The solution for our model
is therefore of the
type encountered in the bosonic models described in
[\Ref{HMN}--\Ref{BNNW}] rather than the fermionic models. To find the first
few terms in the
expansion of the solution one has to write the Fourier transform of the
kernel in the form $1/(G_+(\omega)G_-(\omega))$ where $G_\pm(\omega)$ are
analytic in the upper (lower) half planes with
$G_-(\omega)=G_+(-\omega)$. This determines uniquely
$$\eqalign{
G_+(\omega)=&{\Gamma(\half-i(1-2\Delta)\omega/2)\Gamma(1-i\Delta\omega)\over
\Gamma^2(\half-i\omega/2)}e^{-\half\ln(-i\Delta\omega)}\cr
&\times e^{-i\omega(\half-\Delta)(1-\ln(-i\omega(\half-\Delta)))
-i\omega\Delta(1-\ln(-i\omega\Delta))+i\omega(1-\ln(-i\half\omega))}.\cr}
\efr
Following the discussion in [\Ref{BNNW}], if
$G_+(i\xi)$ has an expansion for small $\xi$ like
$$
G_+(i\xi)={k\over\sqrt\xi}e^{-a\xi\ln\xi}\left(1-b\xi+{\cal
O}(\xi^2)\right),
\nfr{EXG}
then the first few terms of the free-energy for $h\gg m$ are given by
$$\eqalign{
\delta f(h)=&-{h^2k^2\over4}\left[\ln{h\over m}
+\ln\left({\sqrt{2\pi}ke^{-b}\over G_+(i)}\right)-1+
a(\gamma_{\rm E}-1+\ln8)\right.\cr
&\qquad\qquad\qquad\left.
+(a+\half)\ln\ln{h\over m}+
{\cal O}\left({\ln\ln(h/m)\over\ln(h/m)}\right)
\right].\cr}
\efr
Our kernel does indeed have an expansion of the form \EXG\ with
$$
k={1\over\sqrt{\pi\Delta}},\quad
a=-{1\over2},\quad{\sqrt{2\pi}ke^{-b}\over G_+(i)}={\sin(\pi\Delta)
\over\pi\Delta}e^{\gamma_{\rm E}/2+\left(\frac{3}{2}+2\Delta\right)\ln2},
\efr
and so the first few terms in the free-energy are
$$
\delta f(h)=-{h^2\over4\pi\Delta}\left[\ln{h\over m}
+\ln\left({\sin(\pi\Delta)\over\pi\Delta}2^{2\Delta}\right)-{1\over2}+
{\cal O}\left({\ln\ln(h/m)\over\ln(h/m)}\right)
\right].
\nfr{FESM}

\chapter{Comparison and Conclusions}

Comparing \FESM\ with \FEP\ we see that the result from the TBA
calculation correctly produces the universal coefficients of the
beta-function and, furthermore, we extract the value for the mass-gap
of the supersymmetric O($N$) sigma model:
$$
{m\over\Lambda_{\overline{\rm MS}}}=2^{2\Delta}\cdot{\sin(\pi\Delta)\over
\pi\Delta}, \qquad \Delta = {1 \over N-2} ,  \qquad N > 4.
\nfr{MG}
It is interesting to compare this  result with a calculation in the
large $N$ limit. We find from above that
$m/\Lambda_{\overline{\rm MS}}=1+(2\ln2)/N+ {\cal O}(1/N^2)$ which agrees with
the large $N$ analysis in [\Ref{GRAC}]. Unfortunately the conjecture
for the mass-gap for all $N$ in [\Ref{GRAC}] is not correct because it
is based on a mistaken ansatz for the functional dependence on $\Delta$.

Our calculations provide convincing evidence that the S-matrix of
Shankar and Witten does indeed describe the supersymmetric O($N$)
sigma model. In particular, as emphasized in [\Ref{BNNW}], the
addition of CDD factors of the form \CDDF\
to the S-matrix would change the kernel $R(\t)$ and drastically alter the
thermodynamics of the system, including the expression \FESM\ for the
free-energy, consequently destroying the remarkable agreement with the
perturbative result.

\references

\beginref
\Rref{FNW}{P. Forg\'acs, F. Niedermayer and P. Weisz, Nucl. Phys. {\bf
B367} (1991) 123}
\Rref{HN}{P. Hasenfratz and F. Niedermayer, Phys. Lett. {\bf B245}
(1990) 529}
\Rref{BNNW}{J. Balog, S. Naik, F. Niedermayer and P. Weisz, Phys. Rev. Lett.
{\bf69} (1992) 873\newline
S. Naik, Nucl. Phys. {\bf B} (Proc. Suppl.) {\bf30} (1993) 232}
\Rref{HMN}{P. Hasenfratz, M. Maggiore and F. Niedermayer, Phys. Lett.
{\bf B245} (1990) 522}
\Rref{ZZ}{A.B. Zamolodchikov and Al. B. Zamolodchikov, Ann. Phys.
{\bf120} (1979) 253}
\Rref{THIII}{T.J. Hollowood, Phys. Lett. {\bf B329} (1994) 450}
\Rref{TBA}{E.H. Lieb and W. Liniger, Phys. Rev. {\bf130} (1963) 1605\newline
Al.B. Zamolodchikov, Nucl. Phys. {\bf B342} (1990) 695}
\Rref{PW}{A. Polyakov and P.B. Wiegmann, Phys. Lett. {\bf B131} (1983)
121}
\Rref{W}{P.B. Wiegmann, Phys. Lett. {\bf B141} (1984) 217}
\Rref{SMGN}{A.B. Zamolodchikov and Al.B. Zamolodchikov, Nucl. Phys.
{\bf B133} (1978) 525\newline
M. Karowski and H.J. Thun, Nucl. Phys. {\bf B190} (1981)
61}
\Rref{AHN}{C. Ahn, {\sl Thermodynamics and form factors in
supersymmetric integrable theories\/}, Trieste preprint IC/93/144,
{\tt hep-th/9306146}}
\Rref{CGN}{P. Forgacs, S. Naik and F. Niedermayer,
Phys. Lett. {\bf B283} (1992) 282}
\Rref{REN}{L. Alvarez-Gaum\'e, D.Z. Freedman and S.K. Mukhi, Ann.
Phys. {\bf134} (1981) 85}
\Rref{GVZ}{M.T. Grisaru, A.E.M. de Ven and D.
Zanon, Nucl. Phys. {\bf B277}
(1986) 388; 409}
\Rref{SW}{R. Shankar and E. Witten, Phys. Rev. {\bf D17} (1978) 2134}
\Rref{SHOU}{K. Schoutens, Nucl. Phys. {\bf B344} (1990) 665}
\Rref{EH}{J.M. Evans and T.J. Hollowood, {\sl Integrable theories that
are asymptotically CFT\/}, Preprint CERN-TH.7293/94, SWAT/93-94/32,
{\tt hep-th/9407113}}
\Rref{GRAC}{J.A. Gracey, Phys. Lett. {\bf B298} (1993) 116}
\Rref{Wit}{E. Witten, Phys. Rev. {\bf D16} (1977) 2991}
\Rref{EH2}{J.M. Evans and T.J. Hollowood, {\sl The exact mass-gap of
the supersymmetric $\CP$ sigma model\/}, preprint CERN-TH.7426/94,
SWAT/93-94/42}
\Rref{DF}{P. DiVecchia and S. Ferrara, Nucl.~Phys.~{\bf B130} (1977) 93}
\endref
\ciao